\documentclass[runningheads]{llncs}
\usepackage{cite}
\usepackage[T1]{fontenc}
\usepackage[colorinlistoftodos]{todonotes}
\usepackage{amsmath}
\usepackage{amssymb}
\usepackage{cleveref}
\usepackage{array}
\usepackage{makecell}
\usepackage{multirow}
\usepackage{graphicx}
\begin{document}
\title{Online Change-point Detection for Cooperative Multi-Agent Reinforcement Learning}
\titlerunning{Online Change-point
Detection for Cooperative MARL}
%
\author{Fatemeh Saberi Khomami\inst{1}\orcidID{0009-0007-0823-3887} \and
Julita Vassileva\inst{1}\orcidID{0000-0001-5050-3106} }
\authorrunning{F. Saberi Khomami and J. Vassileva}
%
\institute{Department of Computer Science, University of Saskatchewan,
 Saskatoon, Canada \\
\email{f.saberi@usask.ca, jiv@cs.usask.ca}}
\maketitle              
\begin{abstract}
Cooperative multi-agent reinforcement learning (MARL) systems rely on 
past experience for learning coordinated behaviour, but this experience 
may become unreliable if the environment or task objective changes during 
training. In such cases, agents first need a way to recognize that the 
situation has changed before deciding how to adapt. This paper studies 
online change-point detection for cooperative MARL using reward-derived 
signals. We propose \emph{Patterns of Past Rewards} (PPR), a lightweight 
algorithm-agnostic detector that smooths agents' return streams, highlights 
recent changes, and applies a statistical drift detector to flag significant 
shifts. We evaluate PPR in a custom Speaker-Listener environment based 
on the Multi-Agent Particle Environment under two controlled non-stationarity 
scenarios. Our results show a trade-off between detection speed and alarm 
stability. A smoothed-return baseline detects earlier but produces many 
repeated alarms. In contrast, applying the detector directly to raw returns 
often misses the shift. PPR offers a more balanced approach by limiting 
redundant detections while still identifying the controlled shifts. 
These findings highlight PPR as a lightweight, reward-based monitoring 
tool that enables cooperative MARL systems to reliably identify major 
changes during training.

\end{abstract}
\section{Introduction}
Multi-agent reinforcement learning (MARL) has become an important approach for cooperative tasks 
in which multiple agents must coordinate over time, such as warehouse logistics, traffic management, 
and collaborative robotics~\cite{Abdoos2011,Krnjaic2024}. 
In these systems, agents typically acquire policies by repeatedly interacting with an environment and 
with each other, gradually shaping behaviour through accumulated experience~\cite{albrecht2018}. This reliance on 
past experience is acceptable as long as the learning circumstances remain compatible with the conditions 
seen during training. However, real-world multi-agent systems are rarely stationary. Observation channels 
can change, reward functions can be updated, and the composition or roles of agents may 
vary over time~\cite{khetarpal2022}. When such shifts (non-stationarities) occur, 
previously effective policies can become unreliable.

This paper is motivated by a fundamental question: \emph{when should a learning system stop trusting 
what it has learned?} In cooperative MARL, this question matters because agents depend on past 
observations, coordination patterns, and reward histories, which may become invalid after a 
shift~\cite{papoudakis2019}. If the system continues to trust stale experience for too long, 
learning may become inefficient or misleading. If it reacts too aggressively, it may discard 
useful knowledge and destabilize coordination. The challenge is to detect when the environment has 
changed enough so that prior experience should be treated with caution, rather than treating 
non-stationarity as a single abstract property of the process.

In MARL and non-stationary reinforcement learning (RL), shifts are usually addressed through 
adaptation, such as continued training, exploration adjustment, or other forms of policy 
recovery~\cite{zhang2023,steinpuarz2022}. These approaches are useful, but they focus primarily on 
online adaptation mechanisms, rather than treating the detection of non-stationarity itself as a 
central objective. In practice, detection must come first because without an online signal 
of change, an agent has no clear reason to adjust its trust in prior knowledge. Thus, change-point detection 
is a foundational step.

To address this problem, we build on the idea that reward-derived signals implicitly encode information 
about environment dynamics, task objectives, and agent coordination quality, and if a change regarding 
any of these aspects occurs, it should be reflected in the reward signal. We designed 
\emph{Patterns of Past Rewards} (PPR), a lightweight, algorithm-agnostic detector that operates on windows 
of episodic returns without modifying the underlying MARL algorithm. We evaluate 
this approach in a custom Speaker-Listener environment derived from the Multi-Agent Particle Environment 
(MPE)~\cite{mordatch2017,Lowe2017}, where we introduce controlled shifts during training. 
The experiments are designed to assess whether online change-point detection is feasible 
in these controlled cooperative settings. 
The paper's contribution is a step toward trustworthy multi-agent systems under objective and 
environmental shifts, in which systems first detect such shifts and then decide how to adapt.

The remainder of the paper is organized as follows.~\Cref{sec:background} 
reviews cooperative MARL, non-stationarity, and online change-point detection.~\Cref{sec:approach} 
introduces the PPR pipeline.~\Cref{sec:experiment} describes the experimental setup and controlled 
shift scenarios.~\Cref{sec:results} presents the results 
and ablation comparison. Finally,~\Cref{sec:conclusion} summarizes the conclusions of the study.
\section{Background}
\label{sec:background}
\subsection{Cooperative MARL under Partial Observability}

MARL adds flexibility to RL by introducing a multi-agent environment in which agents interact in a shared 
environment over time to adapt to a more complex goal. In fully cooperative settings with partial observability, 
the problem is often modelled as a decentralized partially observable Markov decision process 
(Dec-POMDP)~\cite{Bernstein2002, Oliehoek2012}. This formalism captures the key properties of our setting: 
multiple agents, local observations, and a shared reward signal.

A Dec-POMDP is defined by the tuple $(\mathcal{I}, \mathcal{S}, \mathcal{A}, T, \mathcal{R}, \mathcal{Z}, \mathcal{O}, \gamma)$,
where $\mathcal{I}=\{1,\dots,N\}$ is the set of agents, $\mathcal{S}$ is the state space, 
$\mathcal{A}=\mathcal{A}_1 \times \cdots \times \mathcal{A}_N$ is the joint action space, $T$ is 
the transition function, $\mathcal{R}$ is the shared reward function, $\mathcal{Z}$ is the joint 
observation space, $\mathcal{O}$ is the observation function, and $\gamma \in [0,1)$ is the discount 
factor~\cite{Bernstein2002}. At each time step, the environment is in a state $s_t \in \mathcal{S}$. 
Each agent $i \in \mathcal{I}$ receives a local observation $z_t^i$, selects an action $a_t^i \in \mathcal{A}_i$ according 
to its policy, and the joint action $\mathbf{a}_t$ determines the next state through $T$. The agents then receive a 
shared reward $r_t = \mathcal{R}(s_t, \mathbf{a}_t)$.

The objective in cooperative MARL is to learn policies that maximize the expected discounted 
team return,
\[
J(\pi) = \mathbb{E}\left[\sum_{t=0}^{\infty} \gamma^t r_t \right].
\]
Unlike single-agent RL, each agent acts based only on its local observation history rather than 
the full environment state. As a result, performance depends not only on the 
quality of individual decisions, but also on coordination among agents over time~\cite{albrecht2018,papoudakis2019}. 
In this work, we instantiate our framework in a two-agent cooperative communication task derived from MPE, 
where agents must coordinate through messages and actions to achieve a shared goal~\cite{Lowe2017,mordatch2017}.
\subsection{Non-stationarity in RL and MARL}

Many RL algorithms assume that the transition dynamics and reward function do not change over time. 
In such stationary environments, experience from the past can be a representative of future 
interactions~\cite{khetarpal2022,peng2024}. In practice, this assumption often fails because changes in the environment's 
behaviour, reward structure, or observation process can alter the distribution of trajectories an agent experiences over time.

Numerous studies have introduced approaches for single-agent RL in non-stationary environments. 
Hamadanian et al.\ considered environments whose behaviour depends on a time-varying latent context 
and studied how to update policies online while limiting forgetting of behaviour that remains useful 
across contexts~\cite{Hamadanian2025}.

In multi-agent settings, non-stationarity arises from both changes in the environment and other 
agents simultaneously learning. For any individual agent, changes in teammates' or opponents' policies 
can make the effective transition dynamics non-stationary, even if the environment itself remains 
unchanged~\cite{papoudakis2019}. Papoudakis et al.\ surveyed this phenomenon in deep MARL and reviewed 
techniques such as centralized critics, opponent modelling, and experience replay modifications that 
aim to reduce instability due to changing co-players. More recent work proposed 
multi-timescale learning rules to better track non-stationary behaviour in decentralized cooperative MARL~\cite{nekoei2023}, and fast adaptation schemes for sudden policy changes in teammates~\cite{zhang2023}.

\subsection{Change-Point Detection}

Change-point detection seeks to identify when the distribution generating a sequence of observations has changed. 
In RL, this is important because shifts in dynamics or rewards can make earlier experience less 
reliable~\cite{hadoux2014,canonaco2020,li2025}. 
In cooperative MARL, the problem is amplified because one agent's policy change can alter the observations and 
rewards seen by others~\cite{papoudakis2019,albrecht2018}.

Classical likelihood-based sequential methods, such as Page's CUSUM and Shiryaev-type tests, 
provided a strong foundation for quick change detection. 
~\cite{banerjee2017,hadoux2014}. These methods are effective when a likelihood model is available 
or can be approximated, but they are less directly suited to model-free RL monitoring, where the signal 
is often a reward trace or another summary statistic rather than a fully specified transition model~\cite{canonaco2020,li2025}.
For this reason, model-free detectors have received increasing attention in RL. Canonaco et al. 
combined a statistical hypothesis test with an importance-sampling transformation and a CUSUM-style upper 
layer to detect non-stationarity online, then triggered adaptation by resetting optimizer 
state~\cite{canonaco2020}. More recent work also studied change-point detection in offline 
RL by testing the stationarity of the optimal \(Q\)-function and using the detected change 
point to select the data segment for subsequent learning~\cite{li2025}. These studies support the broader 
view that detection and adaptation are distinct steps: the first determines whether a change 
has occurred, and the second decides how to respond.

In streaming settings, a practical alternative is sliding-window distribution testing. 
KSWIN follows this idea by comparing a reference window and a recent window with a Kolmogorov--Smirnov (KS) test, 
and it is implemented in the River library as an online concept-drift detector with tunable window and significance 
parameters~\cite{raab2020}. This makes it suitable for reward-stream monitoring in RL, where the signal is 
available online but is often noisy. Related RL work also uses KS-based statistical ideas for context detection. 
Dick et al.\ used optimal-transport distances and an adapted KS test to detect and label task changes from online 
experience streams, while De Hauwere et al.\ used a KS test to detect when augmented state information becomes 
necessary in delayed coordination problems~\cite{dick2025,dehauwere2011}. In our setting, change-point detection 
serves as a diagnostic step that indicates when past experience should be treated with caution before any later 
adaptation mechanism is applied.

Reward-based monitoring has also been used in non-stationary RL and traffic-control settings. 
Salkham and Cahill used moving-average rewards as part of an online change-detection mechanism in fluctuating traffic 
control, and later work similarly used reward-based signals to detect environment drift in RL~\cite{salkham2010,fang2024}. 
The key limitation of reward-based monitoring is that reward-derived sequences are often noisy, 
which can reduce detector stability and delay the identification of meaningful shifts.
\section{Our Approach}
\label{sec:approach}
In this section, we describe \emph{Patterns of Past Rewards} (PPR), a lightweight, 
algorithm-agnostic online detector that monitors the episodic return stream during 
cooperative MARL training. The episodic return is a reward-derived signal computed from 
the rewards collected within an episode. PPR does not modify the underlying learning algorithm or its update rules; 
instead, it reads the return signal produced during training, transforms observed episode returns, 
and applies statistical drift detection to identify when recent return patterns differ from previous patterns. 

Let \(R_e\) denote the return observed at the end of episode \(e\). The simple moving average (SMA) 
stage maps \(R_e\) to \(\bar R_e\), the exponential moving variance (EMV) stage maps \(\bar R_e\) to \(V_e\), 
and KSWIN maps \(V_e\) to the binary drift indicator \(d_e \in \{0,1\}\), 
which indicates whether drift is detected after episode \(e\). The overall pipeline is

\[
R_e \xrightarrow{\mathrm{SMA}} \bar R_e \xrightarrow{\mathrm{EMV}} V_e \xrightarrow{\mathrm{KSWIN}} d_e .\]
The SMA stage reduces short-term fluctuations in the return stream, while the EMV stage 
emphasizes changes in the smoothed return trajectory.  KSWIN then tests whether the 
recent transformed return distribution differs from the distribution in the previous window.
\paragraph{Simple Moving Average (SMA).}
The episodic return stream can contain high-frequency fluctuations caused by 
exploration noise, stochastic transitions, and ordinary learning dynamics. 
Applying a drift detector directly to this raw sequence can therefore produce unstable 
alarms that reflect transient variation rather than a persistent change in 
the training regime. PPR first applies SMA to obtain a local 
trend estimate of the recent return pattern. 
Given a window size \(w\), the smoothed return at episode \(e\) is
\[
\bar R_e = \frac{1}{w}\sum_{i=e-w+1}^{e} R_i.
\]
In practice, the moving average can only be computed once at least \(w\) episode 
returns have been observed, so the detector begins after an initial warm-up period.

\paragraph{Exponential Moving Variance (EMV).}
After smoothing the episodic return stream, PPR applies a second transformation 
before statistical drift detection is performed. PPR computes the EMV signal as
\[
V_e = \beta (\bar R_e - \bar R_{e-1})^2 + (1-\beta) V_{e-1},
\]
where \(V_e\) is the EMV value at episode \(e\), and \(\beta \in [0,1]\) controls 
how strongly recent changes influence the updated variance estimate. Larger values 
of \(\beta\) produce a more responsive signal, while smaller values retain more 
past variation and yield smoother behaviour.

This transformation converts changes in the smoothed return sequence into a positive 
variability signal. Gradual changes in \(\bar R_e\) produce relatively small EMV 
values, whereas abrupt changes or bursts of variability produce larger values, making 
sudden changes in return dynamics more distinguishable before the signal is passed to KSWIN.
\paragraph{KSWIN drift test.}
The final stage applies a Kolmogorov--Smirnov sliding-window detector to the transformed values 
\(V_e\). Let \(B\) denote the most recent test window of size \(m\), and let \(A\) denote a 
reference sample drawn from the preceding portion of the KSWIN window (shown by $\mathcal{W}$). 
KSWIN compares their empirical cumulative distribution functions 
\(F_A\) and \(F_B\) using the Kolmogorov--Smirnov statistic
\[
D = \sup_x \left|F_A(x) - F_B(x)\right|.
\]
In our implementation, the critical threshold is computed as
\[
\tau_\alpha = \sqrt{\frac{-\ln(\alpha)}{m}},
\]
where \(\alpha\) is the significance level. A drift is flagged after episode \(e\) when $D > \tau_\alpha$.

\section{Experiment Setup}
\label{sec:experiment}
\subsection{Speaker-Listener Task}

We evaluate PPR in a custom cooperative Speaker-Listener task derived from MPE.
The setting involves two agents with complementary roles: a speaker that observes the goal colour 
and communicates a message vector, and a listener that observes the landmark positions and the speaker's 
message and chooses movement actions. Both agents receive the same team reward, so learning is driven by 
shared task performance rather than by separate local objectives.

In each episode, one landmark colour identifies the goal, and the listener is rewarded according to its 
distance from that landmark. The agents are not given an explicit symbolic mapping from colours to 
landmarks. Instead, they learn a communication convention through reward feedback, with the speaker learning 
to encode the goal colour in its message and the listener learning to interpret the message and move toward 
the corresponding landmark.

\subsection{Controlled Non-stationarity Scenarios}

To study online change-point detection under controlled non-stationarity, 
we introduce two types of shifts during training. In the first scenario, 
the landmark colour assignment is changed during the course of training. 
For example, we reassign the landmarks from red, green, and blue to cyan, 
magenta, and yellow.
The target is still defined through colour, but the colour-to-landmark 
association changes. As a result, the previously learned communication 
convention may no longer produce the correct coordination pattern, and 
the speaker and listener must re-establish a compatible mapping through 
further interaction.

\begin{figure}[!t]
\centering
\includegraphics[width=\textwidth]{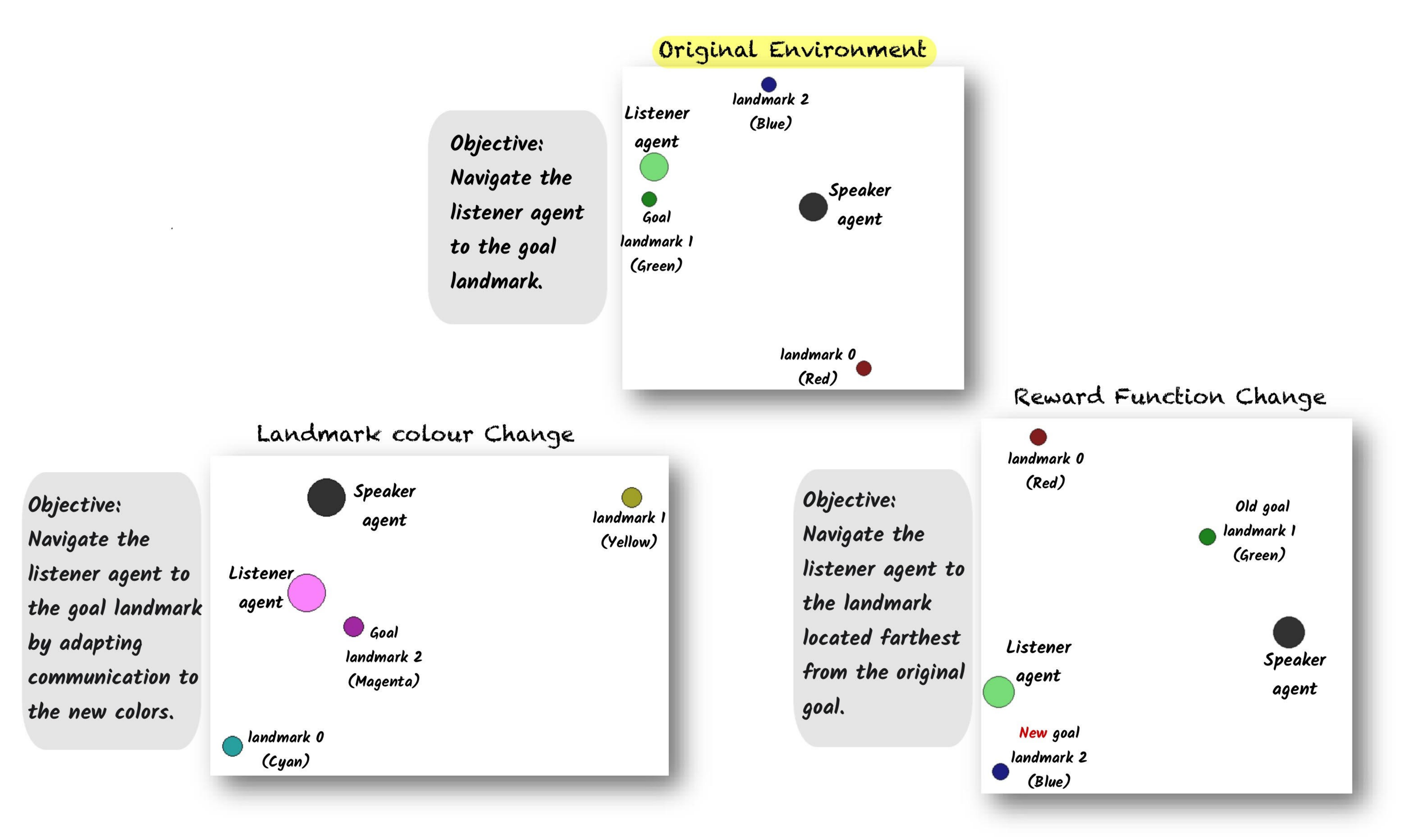}
\caption{Controlled shifts in the custom Speaker--Listener 
environment. The landmark-colour-change scenario reassigns landmark 
colours, while the reward-function-change scenario makes the landmark 
farthest from the original goal the new target.}
\label{fig:custom_env}
\end{figure}

In the second scenario, we change the reward definition while keeping 
the cooperative setting intact. After the shift, the rewarded target 
becomes the landmark farthest from the original goal's location, 
and the team reward is recomputed with respect to this new target. 
Therefore, the agents continue to solve a collaborative task, but the 
communication convention that was useful before the shift may no longer 
match the new reward structure.~\Cref{fig:custom_env} illustrates the original task and the 
two controlled shift scenarios used in our experiments.

\subsection{Training and Detection Protocol}

We train the agents with the Multi-agent deep deterministic policy gradient (MADDPG) algorithm~\cite{Lowe2017} 
in an online setting. PPR runs alongside training and monitors the episodic return stream without 
modifying the MARL algorithm or its update rules. The detector receives the episode-level return 
after each episode, applies its internal transformations, and records a detected change point 
when a shift is flagged.

Each episode contains 25 environment steps. Training is run for \(1,000,000\) steps, 
with the controlled shift introduced at step \(500,000\), corresponding to 
episode \(20,000\). MADDPG uses actor/critic learning rates of \(3\times10^{-4}\) 
and \(10^{-3}\), hidden layers of size \(128\) and \(64\), batch size \(1024\), 
replay buffer size \(10^6\), and \(5000\) warm-up steps.
To assess the contribution of each stage in PPR, we compare the full pipeline, 
\(\mathrm{SMA} \rightarrow \mathrm{EMV} \rightarrow \mathrm{KSWIN}\), against two ablations. 
The first ablation, \(\mathrm{SMA} + \mathrm{KSWIN}\), removes the EMV stage and applies KSWIN 
directly to the smoothed episodic return. The second ablation, \(\mathrm{raw\ return} + \mathrm{KSWIN}\), 
removes both preprocessing stages and applies KSWIN directly to the episodic return stream. For the final comparison, 
all detector variants use the same matched detector setting: \(w=8\), \(\beta=2/3\), \(\alpha=2\times10^{-4}\), 
$\mathcal{W}$ \(=25\), and \(m=10\). The sensitivity of PPR to alternative detector settings is examined in~\Cref{sec:results}.
\subsection{Evaluation Metrics}

We evaluate detection performance using three metrics that capture 
both timeliness and alarm stability. Detection delay is the number of 
episodes between the true shift point and the first detector flag 
after the shift. Pre-shift detections counts the number of flags that 
occur before the true shift point and serves as a measure of false early 
alarms in shift experiments. Excess detections is the number of detector 
flags that occur after the first correct post-shift detection and captures 
the repeated alarm burden after the change has already been identified. 
Together, these metrics distinguish detectors that react quickly 
but over-trigger from detectors that are more conservative but may 
respond later.

A run is counted as a missed detection if the detector does 
not flag any drift after the true shift point within the training horizon. 
In such cases, we report the number of seeds in 
which detection occurred and compute delay only over detected runs.

\section{Results}
\label{sec:results}
\begin{figure}[!t]
\centering
\includegraphics[width=\textwidth]{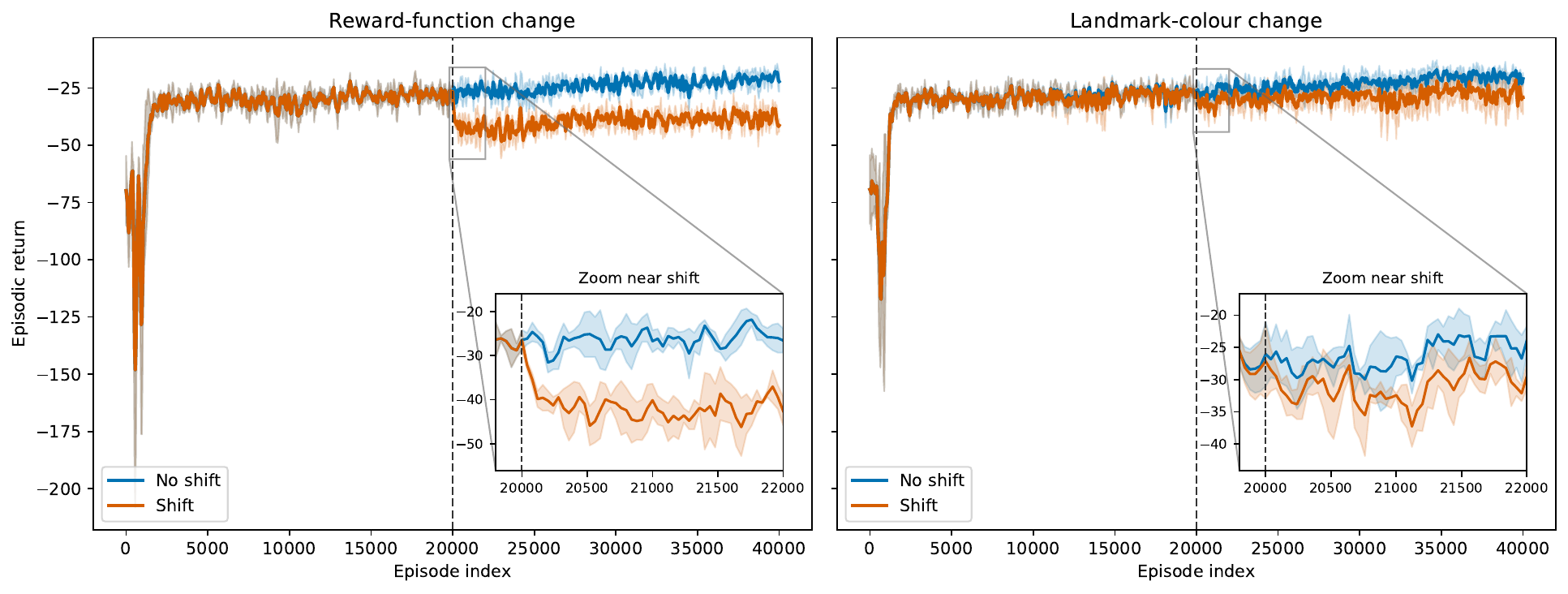}
\caption{Learning curves under controlled shifts compared with matched 
no-shift runs. The dashed vertical line marks the true shift point at episode 20,000.}
\label{fig:learning_curves}
\end{figure}

We first examine the learning dynamics under the two controlled shifts and then evaluate how the different detectors 
respond to these changes.~\Cref{fig:learning_curves} compares shifted runs with matched no-shift runs for both 
scenarios. The dashed vertical line marks the true shift point at episode 20,000. In the reward-function-change panel, 
the shifted curve separates from the no-shift curve shortly after the change point, indicating a visible effect on 
learning performance. In the landmark-colour-change panel, the difference is more subtle, suggesting that some shifts 
do not cause an immediate collapse in average episodic return.

\begin{table}[!t]
\centering
\caption{Detection results for the reward-function-change scenario. Lower values are better for all metrics. Detection delay is reported in episodes. Values are reported as mean \(\pm\) sd over seeds 25--27.}
\label{tab:reward_shift_results}
\begin{tabular}{>{\centering\arraybackslash}p{3.2cm}c@{\hspace{1em}}c@{\hspace{1em}}c}
\hline
Method & Detection delay & Pre-shift detections & Excess detections \\
\Xhline{1.2pt}
PPR & $250.7 \pm 196.1$ & $10.7 \pm 4.6$ & $11.0 \pm 3.0$ \\
\hline
\makecell[c]{SMA\\+\\KSWIN} & $103.7 \pm 76.7$ & $298.3 \pm 6.7$ & $286.3 \pm 17.6$ \\
\hline
\makecell[c]{Raw return\\+\\KSWIN} & \makecell[c]{$18424.0$\\(1/3 detected)} & $0.0 \pm 0.0$ & $0.0 \pm 0.0$ \\
\hline
\end{tabular}
\end{table}

\begin{table}[!t]
\centering
\caption{Detection results for the landmark-colour-change scenario. Lower values are better for all metrics. Detection delay is reported in episodes. Values are reported as mean \(\pm\) sd over seeds 22--24.}
\label{tab:landmark_shift_results}
\begin{tabular}{>{\centering\arraybackslash}p{3.2cm}c@{\hspace{1em}}c@{\hspace{1em}}c}
\hline
Method & Detection delay & Pre-shift detections & Excess detections \\
\Xhline{1.2pt}
PPR & $261.7 \pm 56.6$ & $8.3 \pm 3.1$ & $10.3 \pm 1.5$ \\
\hline
\makecell[c]{SMA\\+\\KSWIN} & $52.0 \pm 48.5$ & $300.3 \pm 14.3$ & $295.7 \pm 12.9$ \\
\hline
\makecell[c]{Raw return\\+\\KSWIN} & \makecell[c]{$4178.0$\\(1/3 detected)} & $0.3 \pm 0.6$ & $0.0 \pm 0.0$ \\
\hline
\end{tabular}
\end{table}
~\Cref{tab:reward_shift_results} shows that SMA + KSWIN detects the reward-function change earlier than PPR, 
but it does so with hundreds of detections before and after the change, which makes the detector difficult to use 
in practice. Raw return + KSWIN is much more conservative: it produces almost no alarms, but it also fails to 
detect the shift reliably. Only one of the three runs triggered detection, and that detection occurred long 
after the true change point. 
PPR sits between these extremes, with slower detection than SMA + KSWIN but substantially fewer repeated alarms 
and better responsiveness than raw return alone.
Detection delay under reward-function change also exhibits high variability across 
seeds, suggesting sensitivity to the underlying MARL training trajectory.

The same pattern appears in~\Cref{tab:landmark_shift_results}. SMA + KSWIN again detects earlier, but its alarm 
stream is highly unstable, with roughly 300 pre-shift and post-shift detections. Raw return + KSWIN remains 
conservative and again misses most shifts. PPR does not minimize delay, yet it provides a more usable balance 
between timeliness and stability, especially when the observed return changes only mildly after the shift.

We further examine the effect of changing parameters \(w\) and \(\beta\) 
around the final detector setting.~\Cref{tab:ppr_sensitivity} shows that reducing 
\(\beta\) gives earlier detection but substantially increases pre-shift detections and 
repeated alarms, while increasing \(w\) reduces alarm burden but increases delay. 
The final configuration was selected as a balanced setting between timely detection 
and alarm stability.
\begin{table}[!t]
\centering
\small
\caption{PPR parameter sensitivity for the landmark-colour-change scenario. KSWIN is fixed at \((\alpha,\mathcal{W},m)=(2\times10^{-4},25,10)\), and values are reported as mean \(\pm\) sd over seeds 22--24.}
\label{tab:ppr_sensitivity}
\begin{tabular}{>{\centering\arraybackslash}p{2.4cm}c@{\hspace{0.6em}}c@{\hspace{0.6em}}c@{\hspace{0.6em}}c@{\hspace{0.6em}}c}
\hline
Setting & \(w\) & \(\beta\) & \makecell[c]{Detection delay} & \makecell[c]{Pre-shift detections} & \makecell[c]{Excess detections} \\
\Xhline{1.2pt}
\textbf{Final} & \textbf{8} & \textbf{0.667} & \(\mathbf{261.7 \pm 56.6}\) & \(\mathbf{8.3 \pm 3.1}\) & \(\mathbf{10.3 \pm 1.5}\) \\
\hline
Smaller \(w\) & 5 & 0.667 & \(674.0 \pm 286.6\) & \(21.3 \pm 1.5\) & \(24.7 \pm 2.5\) \\
Larger \(w\) & 10 & 0.667 & \(1100.3 \pm 697.1\) & \(8.3 \pm 2.5\) & \(7.3 \pm 3.1\) \\
\hline
Smaller \(\beta\) & 8 & 0.500 & \(141.7 \pm 86.7\) & \(44.3 \pm 4.7\) & \(43.0 \pm 2.6\) \\
Larger \(\beta\) & 8 & 1.000 & \makecell[c]{No detection\\(0/3)} & \(0.0 \pm 0.0\) & \(0.0 \pm 0.0\) \\
\hline
\end{tabular}
\end{table}

Across both scenarios, the results indicate a clear sensitivity--stability trade-off. 
A detector that reacts very quickly may also over-trigger and produce many redundant alarms, 
whereas a very conservative detector may avoid false alarms but miss the shift altogether.
Overall,PPR offers a more balanced operating point for online monitoring in cooperative MARL, 
where useful change-point detection must be both sufficiently sensitive and stable.
\section{Conclusion}
\label{sec:conclusion}
This paper examines online change-point detection in cooperative MARL, where shifts in reward structure or 
coordination requirements can make prior experience unreliable. To address this problem, we propose PPR, 
a lightweight detector that monitors episodic returns through a sequence of SMA and EMV transformations, 
followed by a KSWIN drift test. Our method is algorithm-agnostic and can be applied alongside any existing 
MARL training procedures without modifying the underlying learner.

Our experiments in a custom Speaker--Listener environment investigate two types of shifts: a reward-function 
change and a landmark-colour change. Across both scenarios, the results show a clear sensitivity--stability trade-off. 
SMA + KSWIN detects changes earlier but produces a large number of pre-shift and post-shift detections, 
making its alarm stream difficult to use in practice. Raw return + KSWIN is much more conservative, 
but often fails to detect the shift within the training horizon. PPR offers a more balanced outcome by 
reducing repeated alarms while remaining more responsive than raw return monitoring.

These results support the view that reward-derived signals can provide useful diagnostic information about 
non-stationarity in cooperative MARL. At the same time, the observed seed sensitivity, especially under 
changes in the reward function, indicates that reward-based detection remains challenging. 

Future work 
will investigate adaptive detector parameter selection, broader monitoring signals beyond episodic returns, 
and the integration of downstream adaptation mechanisms to enable detected shifts to trigger appropriate 
recovery strategies.

\begin{credits}
\subsubsection{\ackname} This research was supported by the Natural Sciences and Engineering Research 
Council of Canada (NSERC) through a Discovery Grant held by the second author.

\noindent\textbf{Declaration of generative AI in scientific writing}\newline
\noindent During proofreading of this article, ChatGPT (OpenAI) was used to assist in identifying spelling, grammar, and clarity issues. All manuscript text was written and thoroughly reviewed by the authors, who take full responsibility for its content.
\end{credits}
%
%
%
\bibliographystyle{splncs04}
\bibliography{bib}

\end{document}